\documentclass[a4paper,11pt]{article}
\usepackage{jinstpub} 
\usepackage{subcaption}
\usepackage{booktabs}
\usepackage{siunitx}

\title{\boldmath Using Machine Learning to Separate Cherenkov and Scintillation Light in Hybrid Neutrino Detector}

\author{A. Bat}
\affiliation{Bandirma Onyedi Eylul University,\\
10250 Bandirma, Balikesir,Turkey}

\emailAdd{abat@bandirma.edu.tr}

\abstract{This research investigates the separation of Cherenkov and Scintillation light signals within a simulated Water-based Liquid Scintillator (WbLS) detector, utilizing the XGBoost machine learning algorithm. The simulation data were gathered using the Rat-Pac software, which was built on the Geant4 architecture. The use of the WbLS medium has the capability to generate both Scintillation and Cherenkov light inside a single detector.  To show the separation power of these two physics events, we will use the supervised learning approach. The assessment utilized a confusion matrix, classification report, and ROC curve, with the ROC curve indicating a performance result of $0.96 \pm 1.2\times 10^{-4}$. The research also aimed to identify essential parameters for effectively distinguishing these physics events through machine learning. For this, the study also introduced the SHAP methodology, utilizing game theory to assess feature contributions. The findings demonstrated that the number of hits has a significant effect on the trained model, while the mean hit time has a somewhat smaller impact. This research advances the utilization of AI and simulation data for accurate Cherenkov and Scintillation light separation in neutrino detectors.}


\begin{document}
\maketitle
\flushbottom

\section{Introduction}
\label{sec:introduction}

Over the years, water~\cite{Becker-Szendy:1995qpb, Becker-Szendy:1992ufh, FUKUDA2003418, BOGER2000172}, and Liquid Scintillators (LS)~\cite{PhysRevLett.90.021802, PhysRevLett.94.081801, ALIMONTI2009568, Apollonio2003331, PhysRevLett.108.191802, An_2013} have long played an essential role in neutrino as active mediums in detectors. In future experiments, the detector design aims to integrate the benefits of both water and liquid scintillator mediums. The objective is to create an experiment that explores various physics perspectives simultaneously while also achieving cost efficiency in terms of per-unit volume. The Water-based liquid scintillator (WbLS) \cite{YEH201151}, a hybrid detector, has been able to produce both Cherenkov and scintillation light using one target. This feature brings up a wide range of opportunities for particle and neutrino physics studies.


Neutrinos are challenging particles that are difficult to detect due to their limited interaction with matter. In neutrino detectors such as water Cherenkov (WC) and LS detectors, the interaction of neutrinos with the detection medium leads to the production of secondary particles. These particles, as they move faster than light in the medium, emit Cherenkov light in the case of WC detectors, or cause the medium itself to fluoresce, producing scintillation light in LS detectors. Photomultiplier Tubes (PMTs), placed strategically around the detector tank, capture this light. The analysis of the intensity and pattern of the detected light allows scientists to reconstruct the interaction's properties, including the energy and direction of the incoming neutrino.

The concept of WbLS was initially investigated in \cite{4336931}. A wide-ranging study was recently conducted in \cite{YEH201151, 4336931, Land2020oiz, Zhao2023ydx} with the same goal, employing WbLS in massive underground experiments to hunt for unusual physic processes. According to research, WbLS has superb attenuation length, sufficient light yield, and long-term chemical stability all of which are essential for an experiment to succeed~\cite{Bignell_2015}. Furthermore, it also shows a great time response~\cite{D0MA00055H}. These benefits provide physicists the chance to look at both Cherenkov and scintillation physics in a single, affordable experiment.

The Chess experiment~\cite{PhysRevC.95.055801} studies several WbLS formulations and analyzes their properties, providing us with thorough information on light yield and time profiles. The results are compatible with the Rat-Pac software and available for public use~\footnote{https://github.com/AIT-WATCHMAN/rat-pac}. Several experimental studies were carried out and implemented in the Rat-Pac program, yielding results for medium containing 1\%, 3\%, 5\%, and up to 10\% LS as part of the active medium for WbLS. Li et al.\cite{LI2016303} reported on separating scintillation and Cherenkov lights in Linear Alkyl Benzene (LAB), a component of WbLS. They measured the rise and decay times of scintillation light, and the full width of Cherenkov light, which was dominated by the time resolution of the PMTs. Kaptanoglu et al.~\cite{Kaptanoglu2022} described measurements of WbLS, demonstrating the separation of Cherenkov and scintillation components using the fast timing response of a Large Area Picosecond Photodector (LAPPD). This involved characterizing the time profiles of WbLS mixtures with improved sensitivity to the scintillator rise time.

New detectors using the WbLS as an active medium are proposed to expand research on neutrinos. In particular, the proposed THEIA~\cite{Zsoldos:2022uai} 10s-of-kton detector aims to distinguish between scintillation and Cherenkov light in a kilo-ton detector. The THEIA detector will also be able to work on the sun, earth (Geo), and reactor neutrinos. The EOS~\cite{Anderson_2023} experiment is considered a portable detector with approximately four tons (fiducial) and is currently under construction at the University of California-Berkeley laboratories in the USA. The ANNIE experiment conducted at Fermilab deployed WbLS in its experimental setup through the utilization of the SANDI box~\cite{ANNIE:2017nng, ANNIE:2023yny}. The SANDI box is a transparent (acrylic) vessel containing WbLS and weighing 0.5 tons. In 2023, data collection took place for the first time through the integration of WbLS in a large-scale neutrino experiment.


Machine Learning (ML) techniques are extensively used in the field of neutrino physics research to reconstruct the energy and identify and classify the kinds of interactions occurring inside neutrino detectors. This involves arranging the data collected from neutrino detectors, which includes information on particles that generate pixel traces within the detector as a result of their interactions. The particle pictures are then analyzed using computer vision techniques, which help in the detection and distinction of various physical interactions~\cite{Psihas:2020pby}. Deep learning image recognition techniques are the most often used in neutrino experiments. These techniques are frequently used to turn detector data into tensor inputs that resemble pictures.

In the NOvA experiment at Fermilab, the CNN architecture is used to classify neutrino interactions within the detector ~\cite{Aurisano_2016}.  This deep learning CNN architecture was also used in the MicroBooNE experiment~\cite{PhysRevD.103.092003}, which used a liquid argon time projection chamber. Furthermore, the DUNE experiment has created a network that can extract the interaction class and amounts of different particle kinds inside an event. Furthermore, this network can estimate kinematic energy~\cite{PhysRevD.102.092003, DUNE:2020ypp}.

As in many fields of particle physics, Artificial intelligence (AI) algorithms in water Cerenkov detectors can be used to separate signal and background events, make energy estimations, and classify particles. In addition to traditional analysis methods for water Cherenkov and LS detectors, the use of AI-based applications has increased in recent years.
In these detectors, the data can be conceptualized as a chemical trace resulting from the interactions of particles within the detector. These traces are then captured by the detector's PMTs and translated into two-dimensional images. The images are then constructed as tensors and fed into the CNN deep learning architecture. The application of these methodologies can be observed in experiments like Daya Bay~\cite{7838264} and KM3NeT/ORCA~\cite{Aiello_2020}.
Also, the Super Kamiokande experiment employs a feedforward multilayer algorithm based on ROOT's TMLP class to separate the 2.2 MeV gamma signal from background events~\cite{2022.978857}.

Additionally, a comprehensive study was conducted aiming to easily optimize different detector designs~\cite{ELLER2023168011}. Numerous factors can impact the designs of detectors in neutrino and particle physics research, particularly concerning weakly interacting particles like neutrinos. Hence, this article introduces a deep learning model, emphasizing various design parameters for optimizing different detector structures. The neural network (NN) model in this study is a two-step process trained on sensor-based information, proving valuable in the early phases of detector design by enabling the examination of multiple parameters. The model's performance is assessed across different WbLS mediums utilized in hybrid neutrino detectors. The reported results indicate that increasing the scintillation rate leads to improved vertex and energy resolution but decreased directional resolution. These outcomes are considered valuable contributions to enhancing Cherenkov and scintillation separation.

In this research, we conducted the simulation of a detector filled with WbLS using the Geant4-based Rat-Pac simulation tool. The simulation process generated Cherenkov and scintillation physics signals, enabling the differentiation between these two distinct forms of light. Subsequently, artificial intelligence (AI) models were employed to detect and discriminate between Cherenkov and scintillation light signals. To achieve this goal, the AI models underwent training through supervised learning methods, utilizing datasets derived from the simulation. The primary aim of this work is to examine the factors that are essential in successfully distinguishing between Cherenkov and Scintillation signals via the use of machine learning algorithms, which are based on simulation data. The study results demonstrate the algorithm's efficacy and results within the given environment.

\section{Simulation Studies}

Simulation studies were carried out using the Rat-Pac program to distinguish between Cherenkov and Scintillation light signals in the detector loaded with WbLS. Rat-Pac~\footnote{https://github.com/AIT-WATCHMAN/rat-pac} is a comprehensive simulation and analysis package originally developed by S. Seibert. It is built based on the GEANT4, Root, and C++, and utilizes the GLG4Sim libraries~\cite{GEANT4:2002zbu}. This tool combines simulation and analysis into a single framework, simplifying the simulation process. With Rat-Pac, users can easily customize various aspects such as geometry elements, physics details, data structure, and analysis tools to suit the specific requirements of their experiment. One of the key reasons for selecting the Rat-Pac program was its ability to accommodate the WbLS detection environment, including different LS percentages and various isotope additives like Gadalonium. Due to these advantages, Rat-Pac is a favored choice in many neutrino experiments, and its ongoing development ensures continuous improvement and support.

The design implemented using the Rat-Pac program involves a detector with the following dimensions: a cylindrical tank standing at 3 meters tall and having a diameter of 3 meters. The choice of this cylindrical structure is inspired by its widespread use in various currently proposed and employed neutrino detectors. This detector is equipped with 174 Hamamatsu R5912 high quantum efficiency (HQE) PMTs. The used PMT is encapsulated within an 8-inch spherical envelope composed of borosilicate glass and features a bialkali-material photocathode. This PMT is engineered with a box-and-grid configuration of dynodes, comprising ten stages that culminate in a typical amplification factor of $10^7$. As specified by the manufacturer, the quantum efficiency of this device is approximately 23\% at a wavelength of 400 nm, and it possesses an active photocathode surface area of approximately 530 $cm^2$~\cite{AMAUDRUZ2019373}. The selection of these advanced PMTs is motivated by their capability to effectively detect time differences between Cherenkov/Scintillation events. To serve as the detection medium, a mixture of WbLS will be employed, comprising 5\% pure LS and 95\% water. The WbLS, composed of linear alkylbenzene (LAB) as the primary scintillating solvent, is enhanced by the addition of the fluorescent solute 2,5-diphenyloxazole (PPO).  The light yield parameter in the 5\% WbLS medium is measured to be $770 \pm 72$ photons per MeV. The photoluminescence response has the following temporal characteristics: an initial rising time of $0.06 \pm 0.11$ ns, a decay process with a short decay constant of $2.35 \pm 0.13$ ns, and a long decay time constant also recorded at $23.21 \pm 3.28$~\cite{Caravaca2020}.

\begin{figure}[t!]
    \centering
    \includegraphics[ width=0.50\linewidth]{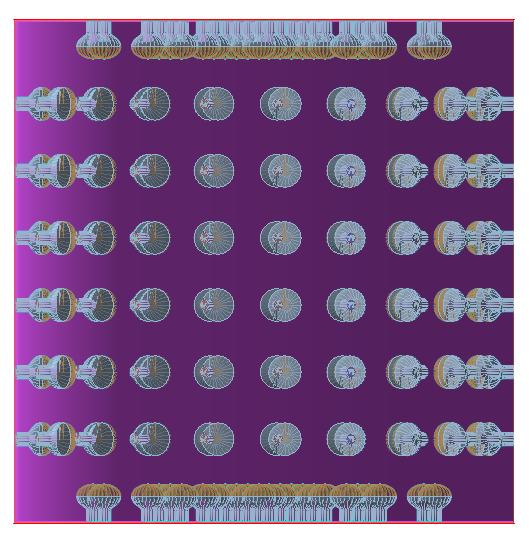}
    \caption{Visualization of the detector and its inner PMTs using the RAT-PAC simulation framework. The detector is 3 meters in height and width. The WbLS medium contains 5
    \% LS and 95\% water. 174 PMTs are placed on detector walls.}
    \label{fig:dt}
\end{figure}

The planned data training method is supervised learning, which depends on labeled datasets to train the model efficiently. The Rat-Pac program will simulate Cherenkov and scintillation light separately, allowing the data to be labeled before the training process begins. Rat-Pac macros will be used during simulations with Cherenkov light-activated and scintillation off to differentiate interactions involving Cherenkov light. To identify interactions caused by scintillation, extra macros will be produced with Cherenkov inactivated and scintillation activated. This method allows each data entry to be labeled individually, making it easier to use the data in the following phases of the AI model.

Rat-Pac primarily employs the GLG4Sim package developed by Glenn Horton-Smith  within GEANT4. This package is utilized to simulate the scintillation process, enhancing the original Geant4 program. The computation of scintillation photon counts is performed as a separate step after all other physics operations have been completed. The GLG4Scint uses the G4OpWLS procedure to keep track of primary scintillations. Then, as well, the G4Cherenkov package inherits from Geaat4 to generate Cherenkov light products.

Neutrinos are typically detected indirectly by searching for the particles produced during their interactions. These interactions are managed by the weak force, which is the fundamental force responsible for neutrino interactions with matter.  One approach to detecting neutrinos is by utilizing their interactions with matter. When neutrinos interact with matter, they can generate detectable signals and particles, which can be measured and analyzed. Indeed, one of the particles that emerges from neutrino interactions is the neutron. Unlike electrons, which produce direct Cherenkov and scintillation light as a result of their interaction with neutrinos, the use of neutrons has several advantages.
Our research focuses on utilizing neutrons produced by neutrino interactions to improve signal differentiation from background noise based on their mass, charge, and interaction characteristics. Neutrinos can be distinguished from background noise by analyzing their timing and spatial distribution of light, which improves the precision of studying neutrino-nucleus cross-sections. Neutron interactions play a crucial role in calibrating neutrino detectors, enhancing the ability of multi-purpose detectors to differentiate between different particle types and interactions. The simulation of neutron generation is a valuable tool in evaluating the performance of detectors used in neutrino experiments. Neutrons generated during these interactions can provide crucial information about the neutrino event, as they are more easily detectable and measurable than neutrinos themselves.

Rat-Pac software uses a single particle gun to generate particles within the detector. This tool generates neutron particles that mimic the original particle's characteristics within the designed detector. The neutron particle's initial momentum was established at zero, accompanied by isotropically random initial directions. The energy of the generated neutron particles ranges from 0.001 MeV to 0.20 MeV, with particle energies selected at random from this range. A total of one million events were simulated for both the Chenkov and Scintillation datasets. Before merging these two datasets, the labels were included in the dataset. Instances indicated as "1" were associated with Cherenkov data, whereas instances labeled as "0" were associated with Scintillation data. To reduce detector noise, instances with less than five photoelectrons were removed.

\begin{figure}[h!]
  \subcaptionbox*{(a) Mean Hit Time (ns)}[.5\linewidth]{%
    \includegraphics[width=0.9\linewidth]{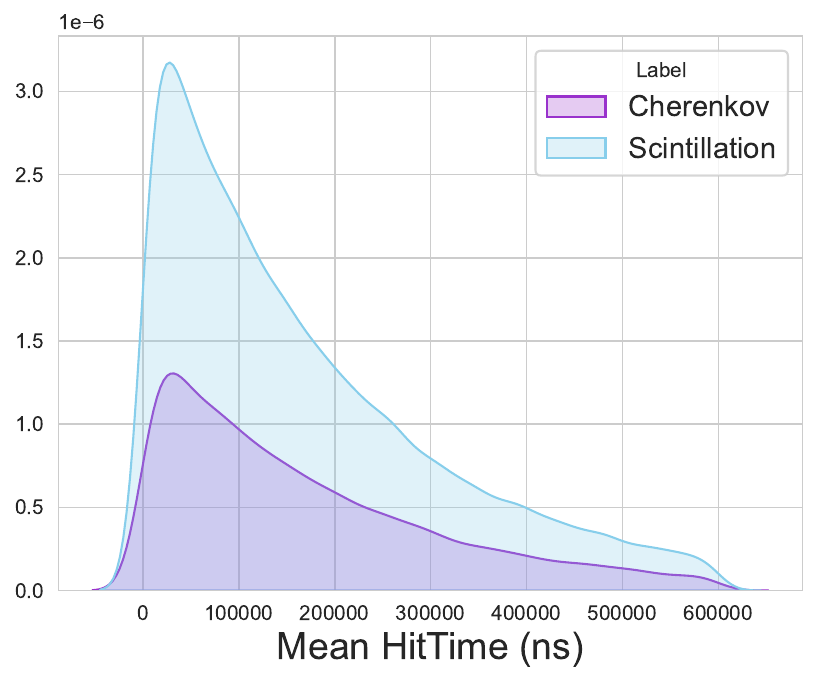}%
  }%
  \hfill
  \subcaptionbox*{(b) Number of Photoelectron}[.5\linewidth]{%
    \includegraphics[width=0.9\linewidth]{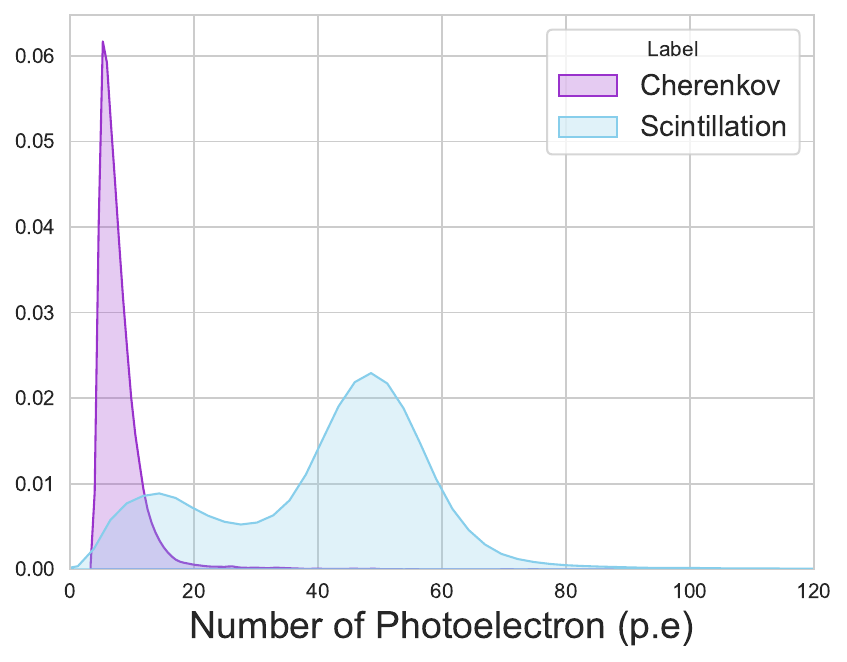}%
  }

  \subcaptionbox*{(c) Sum of charges}[.5\linewidth]{%
    \includegraphics[ width=0.9\linewidth]{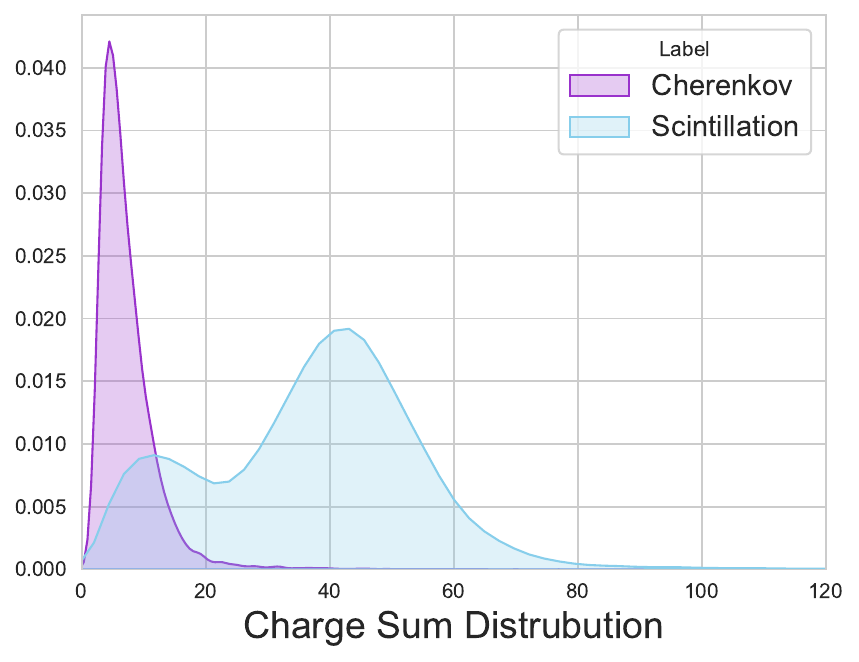}%
  }%
  \hfill
  \subcaptionbox*{(d) Hit position x (mm) }[.5\linewidth]{%
    \includegraphics[ width=0.9\linewidth]{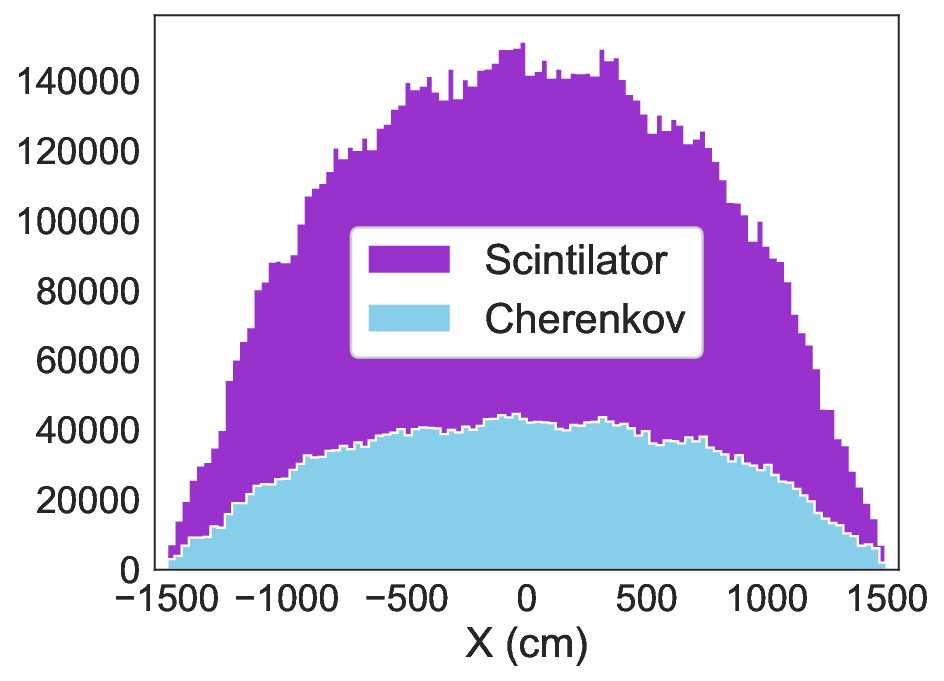}%
  }

   \subcaptionbox*{(e) Hit position y (mm)}[.5\linewidth]{%
    \includegraphics[width=0.9\linewidth]{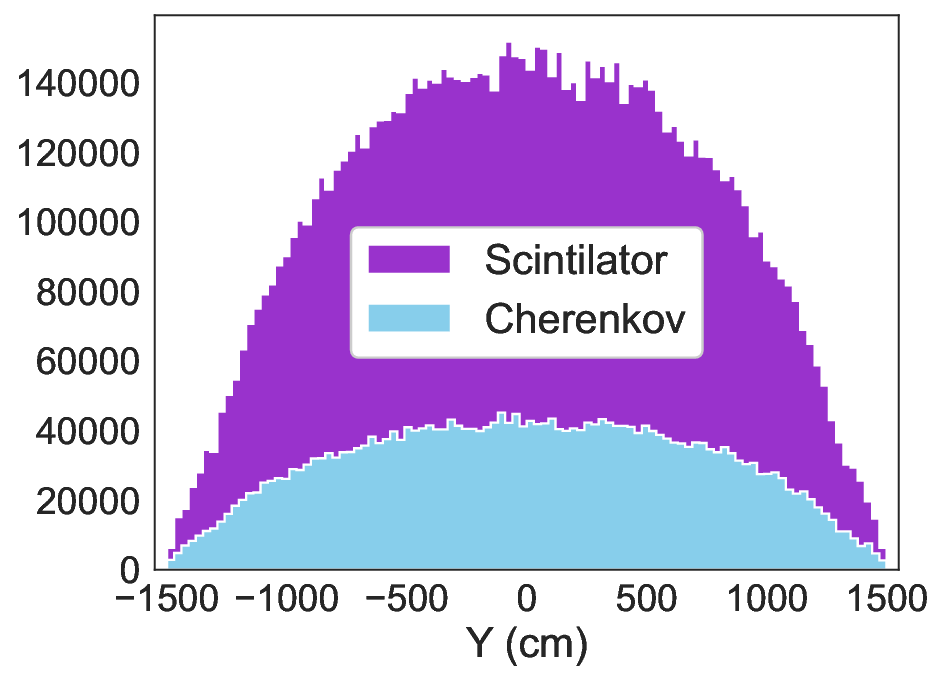}%
  }%
  \hfill
  \subcaptionbox*{(f) Hit position z (mm)}[.5\linewidth]{%
    \includegraphics[width=0.9\linewidth]{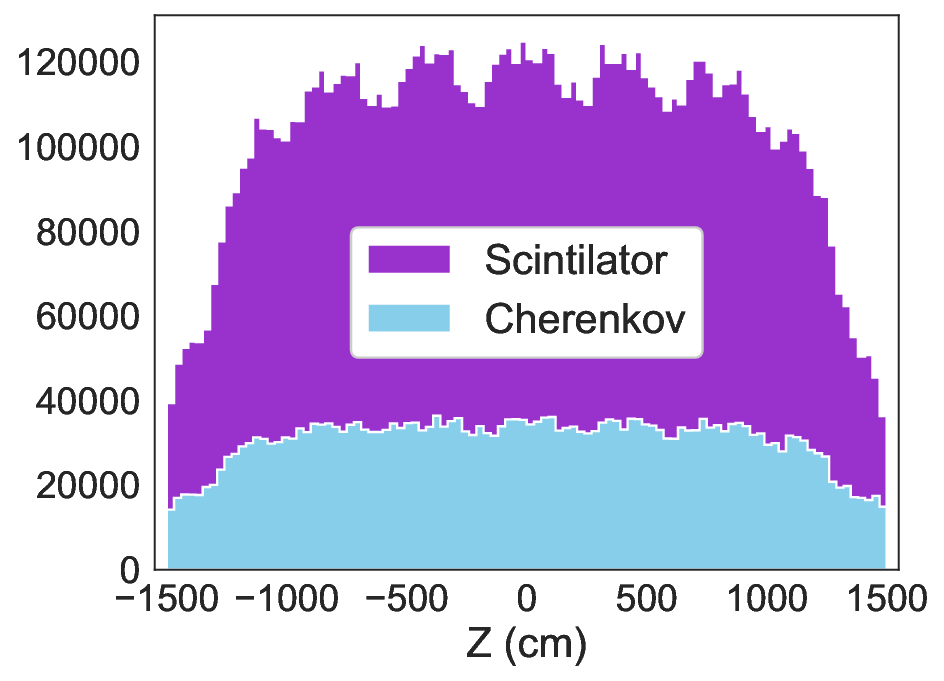}%
  }
  \caption{The figure shows the outputs obtained as a result of the simulation. In the figures, the Scintillation dataset's outcomes are depicted in purple, whereas the Cherenkov dataset's results are represented in blue.}
   \label{fig:features6}
\end{figure}

Figure~\ref{fig:features6} illustrates the outcomes obtained through the simulation. The graphics illustrate the findings of the Scintillation dataset in purple and the results of the Cherenkov dataset in blue. The selected parameters for the ML method are the number of hits (numPE), which refers to the number of hits collected by PMT for an event, mean hit time, the total charge, and the spatial coordinates (x, y, z). The use of the charge and time-based characteristics of observed photons may serve as input useful features for the ML algorithm. The parameters in question serve to collect relevant info on the interactions between particles inside the detector. This includes information about the energy deposited by the particles and the patterns seen in their timing. The use of information associated with the geometry and structure of the detector, including the locations and orientations of the PMTs, has the potential to improve the performance of machine learning algorithms. This parameter plays a crucial role in understanding the response of the detector while improving the process of event reconstruction. These characteristics have the potential to facilitate the differentiation of various particle kinds, including electron, muon, or neutrino events.

\section{Machine Learning Methods}
\label{sec:ml}
This research employed a supervised learning technique to classify the labeled dataset.  The classification task utilized the XGBoost algorithm, which is clarified, along with the chosen parameters, in the subsequent section. The approach involved binary classification, aiming to categorize simulated Cherenkov and Scintillation data. The results were visually represented through a confusion matrix, providing insight into the predictive capabilities. Additionally, the receiver operating characteristic (ROC) plot functioned as a measure of the model's effectiveness.

\subsection{XGBoost Classification}
\label{sec:xgboost}
XGBoost, also known as "\textit{eXtreme Gradient boosting}", is a machine learning algorithm that operates inside the framework of Gradient Boosting Decision Trees (GBDT) \cite{Chen:2016:XST:2939672.2939785}. It provides various favorable additions compared to traditional GBDT methods. The utilization of L1 and L2 regularization approaches is employed to minimize the issue of overfitting, resulting in improved performance and rapid training. By utilizing an ensemble learning methodology, XGBoost integrates numerous weak models in order to provide predictions that are more robust and accurate.

The XGBoost method has received much attention because of its outstanding performance, especially in jobs involving binary classification. The study demonstrates that it has superior performance compared to deep neural network approaches \cite{10.1007/978-3-030-80421-3_37}, especially in the context of binary classification and tabular data. Significantly, this approach demonstrates the multiple accomplishments of achieving high accuracy and delivering fast outcomes. Therefore, the selection of the XGBoost classifier for this study is justified based on these argumentative features.

The XGBoost method has been developed with specific parameters in order to optimize efficiency and limit the chance of overfitting. The features were standardization ahead of the training of the model. In this process, the feature vectors perform a transformation where the average value is subtracted and the units are rescaled to the variation. The Scikit Learn (1.2.1) library was utilized for this purpose. Initially, the dataset is divided into two distinct groups, namely the training set and the validation set. Subsequently, the training dataset was further divided into a test set and a training set. Consequently, the dataset is divided into three subsets for the purpose of model evaluation: 10\% for validation, 18\% for testing, and 72\% for training. The utilization of validation sets serves the purpose of decreasing the risk of overfitting the data. Furthermore, a test model has been employed to evaluate the performance of the model using previously unexplored data. 

The implemented model utilized \textit{n\_estimator} throughout the training process, which may be viewed as the number of epochs in a neural network. The parameters in the model correspond to the representation of the tree, enabling the model to learn iteratively from each tree and afterward modify errors. Therefore, this parameter has an impact on the training of the model. The value of the \textit{n\_estimator} parameter is set to 500. However, if this value is very high, it may lead to overfitting, hence decreasing the effectiveness of the \textit{early\_stopping\_rounds} parameter, which has been set to 20. Utilizing this option will terminate the training process if the model becomes overfit. The tree is limited to a maximum depth of 4. The current configuration shows similarity to the presence of four hidden layers within a basic neural network. To manage the complexity of the model, the values of \textit{min\_child\_weight}, \textit{gamma}, and \textit{alpha} were set to 10, 0.8, and 0.8, respectively. \textit{Alpha} refers to the function responsible for implementing L1 regularization on the weights of the leaf nodes. Additionally, including random elements in the training process can enhance the robustness of the model against noise. To achieve this result, the \textit{subsample} and \textit{colsample\_bytree} parameters are each assigned a value of 0.9. In order to minimize the issue of overfitting, the learning rate is assigned a value of 0.01, which is rather tiny. After each step, the weight for each feature is obtained, and this parameter reduces the feature weights to enhance the conservative nature of the process. The \textit{objective} parameter is set as \textit{binary: logistic}, indicating the use of logistic regression for binary classification and generating the class probabilities as the outcome. This strategy was used to get the optimal set of parameters for this binary classification problem. After determining the optimum parameter set, we utilized all data for training and testing to assess the results. We blend data and obtain the result numerous times to test the model's stability. This will be explained in section~\ref{sec:xgboost_result}.

\section{Results}
\label{sec:result}
\subsection{XGBoost Model Result}
\label{sec:xgboost_result}
\begin{figure}[!h]
     \centering
     \begin{subfigure}[b]{0.45\textwidth}
         \centering
         \includegraphics[width=\textwidth]{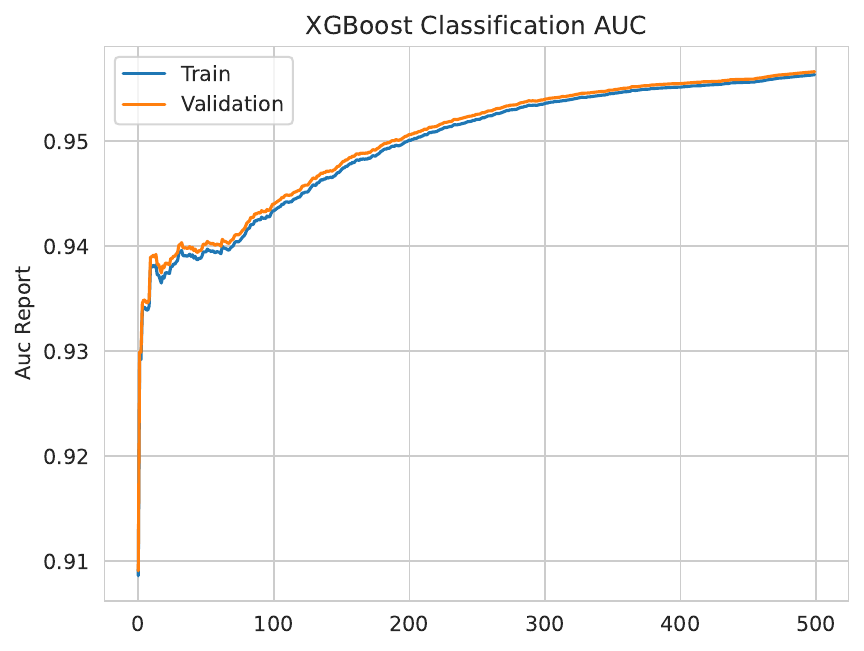}
         \caption{XGBoost Accuracy Report}
         \label{fig:auc}
     \end{subfigure}
     \hfill
     \begin{subfigure}[b]{0.45\textwidth}
         \centering
         \includegraphics[width=\textwidth]{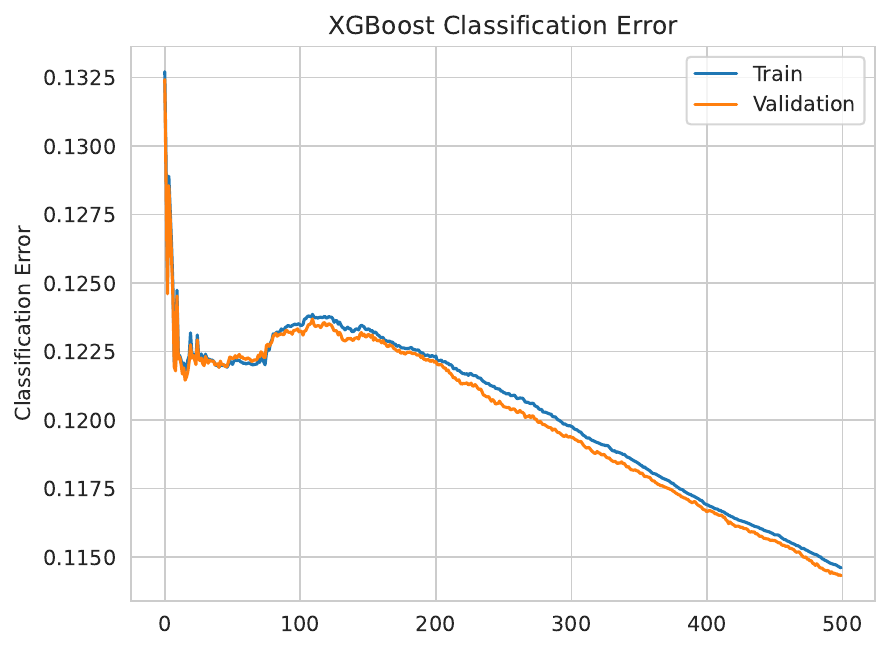}
         \caption{XGBoost Error Report}
         \label{fig:error}
     \end{subfigure}
     \hfill
     \begin{subfigure}[b]{0.45\textwidth}
         \centering
         \includegraphics[width=\textwidth]{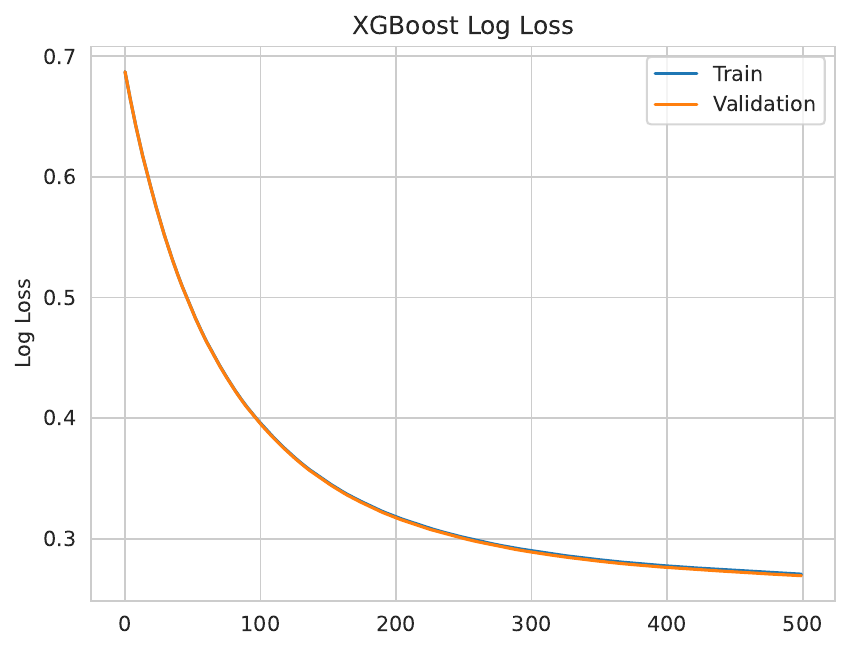}
         \caption{XGBoost Log Loss Report}
         \label{fig:logLoss}
     \end{subfigure}
        \caption{The plots show the evolution of the classification algorithm through iteration based on the training and validation data.}
        \label{fig:progress}
\end{figure}

The model's output was interpreted by the inclusion of the Confusion Matrix (CM), classification report, and ROC curve. The use of a CM allows for the evaluation and comparison of true positives (TP), false positives (FP), false negatives (FN), and true negatives (TN) results. Additionally, the classification report presents the outcomes pertaining to precision, recall, and f1-score metrics. The ROC curve is commonly used to evaluate the performance of classification algorithms. The true positive rate (TPR) is graphed on the vertical axis of the ROC curve, whereas the false positive rate (FPR) is graphed on the horizontal axis of the curve. Figure~\ref{fig:progress} illustrates how the classification algorithm has developed over time through iteration, taking into account both the training data and the validation data. The provided summary plots additionally demonstrate that the training process does not show overfitting throughout the iteration. As expected, the log loss  and the classification error decreases as the accuracy of the model improves with each iteration.

\begin{figure}[t!]
    \centering
    \includegraphics[ width=0.50\linewidth]{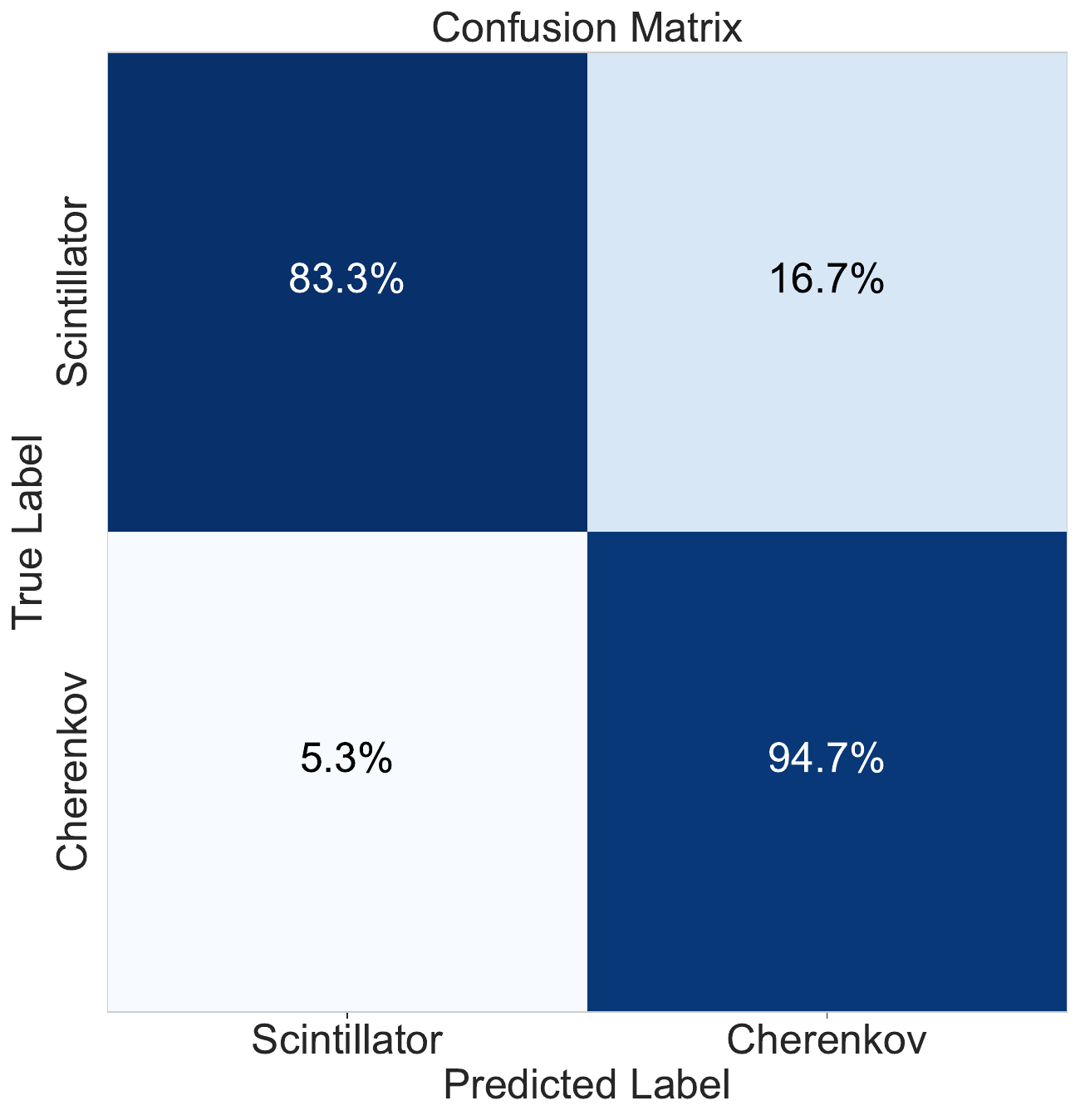}
    \caption{The figure shows XGBoost's confusion matrix. The top row predicts entire Scintillation events (negative case) and the bottom row Cherenkov events (positive case). The x-axis represents model training predictions, and the y-axis represents the original dataset's true cases. The top left is true negative, the top right is false negative, the bottom left is false positive, and the bottom right is true positive.}
    \label{fig:CM}
\end{figure}

In the classification procedure, class 1 is referred to as the positive case. The Cherenkov event was assigned the numerical label of 1, whereas the Scintillator event was assigned the numerical label of 0. The positive situation is represented by the Cherenkov event, whereas the negative case is represented by the Scintillator event. The abbreviation TP can be denoted as Cherenkov events that are accurately foreseen as Cherenkov events. Similarly, the occurrence of scintillator events in TN is expected as an outcome of scintillator events. The figure~\ref{fig:CM} illustrates the confusion matrix for the trained XGBoost model. The abbreviation TN, shown in the top left corner, represents true negative instances of scintillation events that are accurately classified as true. These instances account for 83.3\% of all correctly labeled scintillation events. The term FN (shown in the upper right corner) refers to instances where true scintillator events are mistakenly classified as false. This indicates that around 16.7\% of the events are mistakenly labeled as Cherenkov events, while in reality, they are scintillation events.  Likewise, the bottom part of the visualization shows the outcomes of the Cherenkov incidents. The lower right corner of the figure corresponds to the representation of TP, which denotes the true Cherenkov events (positive cases) that were accurately predicted. The model successfully predicted 94.7\% of the events correctly classified as Cherenkov events. A total of 5.3\% of the Cherenkov events were mistakenly classified as scintillation events, denoted as FP in the bottom left corner. In conclusion, the model has a higher accuracy in properly identifying the Cherenkov events compared to the scintillation events.

The accuracy, recall, and f1-score measures obtained from the classification report are presented in Table~\ref{tab:result1}. The table distinctly illustrates the outcomes for training, testing, and validation, ensuring that the results are not indicative of overfitting. Moreover, the data is evaluated independently for scintillation and Cherenkov events. Subsequently, the upcoming discussion will focus on the Cherenkov scenario, which is a positive case. Precision, calculated as the ratio of correctly identified true positive cases to the total number of positive cases predicted by the model (TP/TP+FP), effectively demonstrates the model's ability to precisely anticipate instances of Cherenkov events that have been categorized as such. The model's predictions provide a precision of Cherenkov with a success rate of  $0.95\pm 7.38\times 10^{-4}$ . Likewise, it was noted that Scintillation had an accuracy of $0.83\pm 1.57\times 10^{-3}$.   Recall, defined as the proportion of TP cases to the actual events (TP/TP+FN), offers insights into how well the model correctly identifies the Cherenkov event within all true Cherenkov events. Based on the recall results, the model achieved valid predictions for $0.95\pm 9.26\times 10^{-4}$  of all Scintillation events and $0.83\pm 1.5\times 10^{-3}$ of all Cherenkov events. The f1-score, which is the harmonic mean of two parameters, serves as a suitable metric for the model performance.

\begin{table}[]
    \centering
    \caption{Results of the classification report for the XGBoost model. The results were obtained for train, test, and validation datasets separately.}
    \begin{tabular}{c c c c c}  \toprule
         &   &    precision & recall & f1-score \\
    \midrule
    Test & Scintilator &   $0.828\pm 1.57\times 10^{-3}$ &    $0.947\pm 9.26\times 10^{-4}$  &    $0.883\pm 1.13\times 10^{-3}$ \\
          & Cherenkov   &   $0.948\pm 7.38\times 10^{-4}$  &   $0.832\pm 1.5\times 10^{-3}$    &    $0.887\pm 1.02\times 10^{-3}$  \\   
    \midrule
    Train & Scintilator &   $0.829\pm 3.05\times 10^{-4}$ &    $0.947\pm 1.32\times 10^{-4}$  &    $0.883\pm 1.65\times 10^{-4}$ \\
          & Cherenkov   &   $0.949\pm 1.4\times 10^{-4}$  &   $0.833\pm 3.13\times 10^{-4}$    &    $0.887\pm 2.8\times 10^{-2}$  \\
      \midrule
     Validation & Scintilator & $0.829 \pm 1.57\times 10^{-3}$ &    $0.946\pm 9.1\times 10^{-4}$  &    $0.88\pm 8.69\times 10^{-4}$ \\
                 & Cherenkov   &   $0.948\pm 1\times 10^{-3}$  &   $0.834\pm 1.4\times 10^{-3}$    &    $0.887\pm 8.8\times 10^{-4}$  \\
      \midrule
     
    \end{tabular}
    
    \label{tab:result1}
\end{table}

\begin{figure}[!ht]
    \centering
    \includegraphics[width=0.80\linewidth]{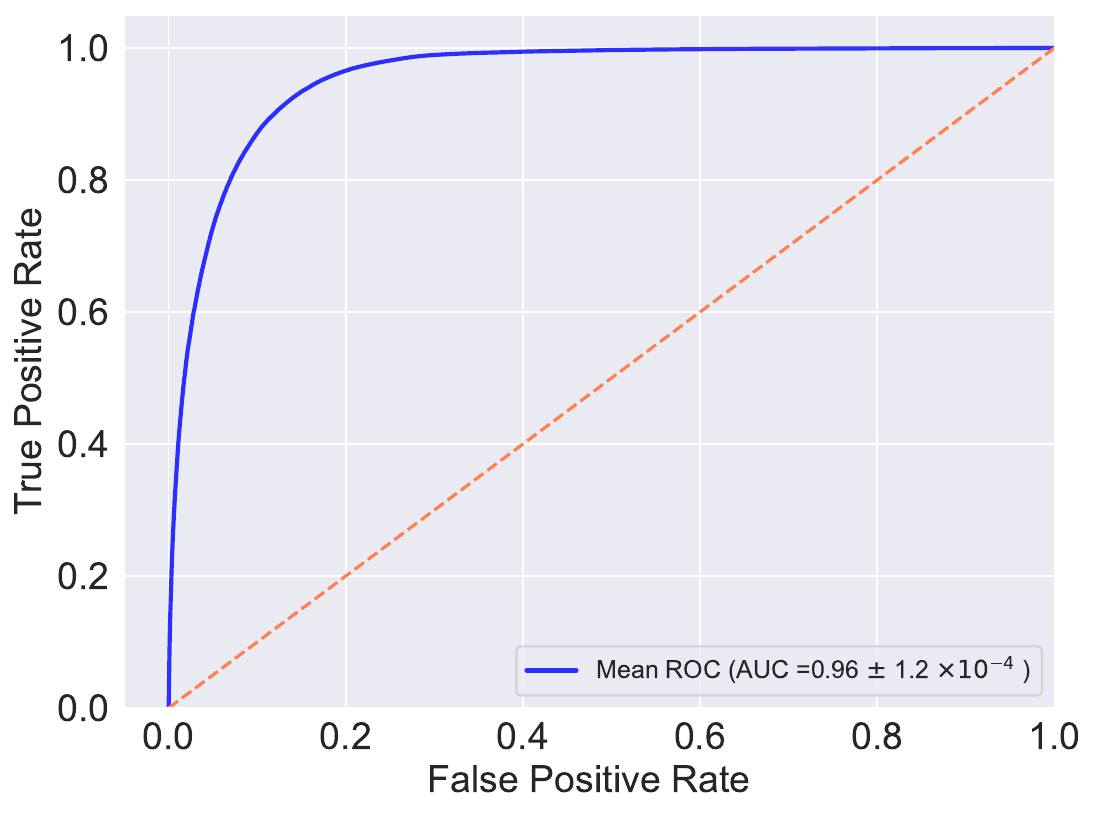}
    \caption{The Receiver Operating Characteristic (ROC) curve displays the True Positive Rate (TPR) on the vertical axis, while the False Positive Rate (FPR) is shown on the horizontal axis of the curve. The Roc accuracy result was acquired using several test and training datasets. The mean ROC value is the average of these ten distinct ROC accuracy results. }
    \label{fig:roc}
\end{figure}

The given figure~\ref{fig:roc} displays the ROC curve for the XGBoost classification methods. The TP rate is shown on the vertical axis of the ROC curve, whereas the FP rate is depicted on the horizontal axis of the curve. ROC curve shows a binary classification success rate. Classifiers that produce curves positioned closer to the upper-left corner demonstrate more precise results. The ROC accuracy estimation was performed using the \textit{predict\_proba} classifier from the Scikit-learn package. This function is used to predict the likelihood of a sample being tagged with a certain class in a classification task. The findings indicate that the ROC  accuracy value for the model is $0.96 \pm 1.2\times 10^{-4}$.

The region bounded by the curve represents the Area Under the Curve (AUC) value. The AUC score evaluates the discriminatory abilities of an algorithm in classification tasks, whereas the ROC curve represents the probability distribution of the algorithm's predictions. In evaluating the AUC, the accuracy score metric from the Scikit-learn package is used. This measure requires the predicted labels for a given sample to precisely correspond to the real class labels. The resultant AUC score is $0.86 \pm 2.5 \times 10^{-4}$. Additionally, the model is run several times to ensure that the predictions are statistically stable. In order to do this, the data was shuffled and distinct test and train samples were used for each iteration.  The aforementioned approach is also used while computing the parameters of the classification report. This is how the standard deviation result is obtained from different runs. The results indicate that the model's outcome has a notable level of stability.

 \subsection{Feature Importance Result}
\label{sec:shapley}

\begin{figure}[b!]
    \centering
    \includegraphics[ width=1\linewidth]{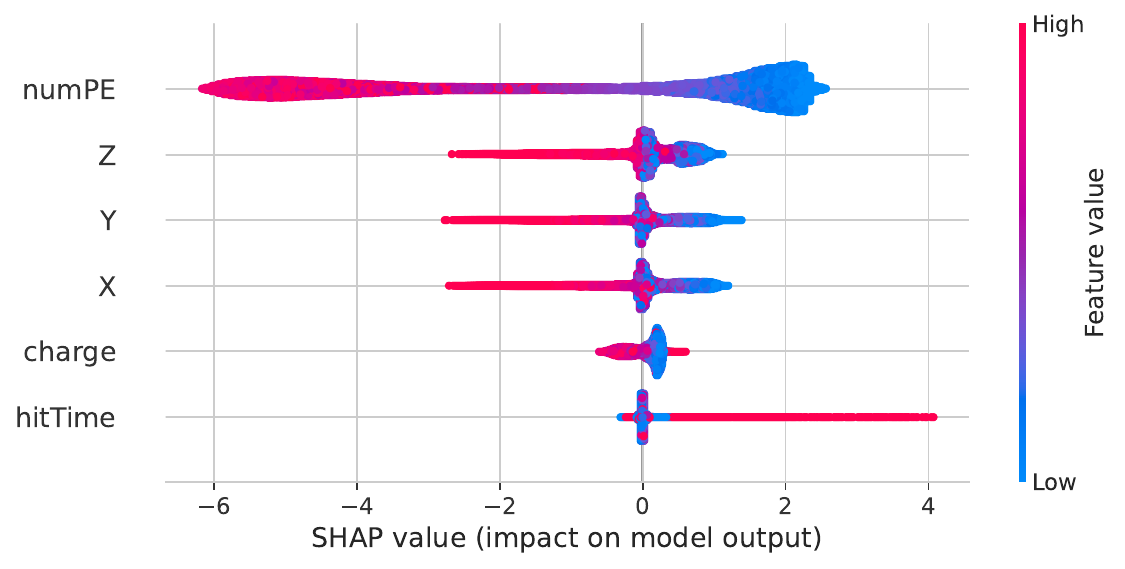}
    \caption{Figure demonstrates the SHAP value for each feature and every event as a dot. The x-axis is the SHAP value and the color bar reflects the feature value (blue is low, red is high). High SHAP values cause the model output to be 1 while low SHAP values cause it to be 0.}
    \label{fig:shap values}
\end{figure}
The SHapley Additive exPlanations (SHAP)~\cite{WINTER20022025} is a methodology that uses a game theory framework~\cite{Shapley1953StochasticG} to assess the individual contributions of each feature towards the ultimate conclusion. Figure~\ref{fig:shap values} shows a `beeswarm' plot which visually represents the distribution and intensity of SHAP values for each feature. Features that exhibit positive SHAP values have a positive influence on the prediction outcome, whereas features with negative SHAP values have a negative influence. The graph shows the SHAP values associated with each feature, which are shown on the vertical y-axis.  Each point on the graph represents an individual entry from the dataset. If there is a high concentration of SHAP values, they are vertically piled up. The figure utilizes color coding to visually show the values of the features, with the color bar serving as a representation of these values. The features with higher values are represented as red, while the features with lower values are represented as blue. Features are also organized along the vertical axis based on their importance, with the most significant features (as determined by the average SHAP value) positioned at the top, and the least significant features positioned at the bottom. According to the figures, the variable with the most impact for this trained model is the number of hits, and the feature with the least impact is the mean hit time. In events when there is a high value of a number of hits, it may be seen that the SHAP value consistently exhibits a decreasing trend. Events with a lower number of hits tend to exhibit high SHAP values, indicating that the model's output value tends towards a higher value of 1. The aforementioned trend is also seen in the characteristics related to charges. The term hit time, often referred to as "arrival time," is associated with a higher value that indicates positive Shap values.

\begin{figure}[!h]
     \centering
     \begin{subfigure}[b]{0.3\textwidth}
         \centering
         \includegraphics[width=\textwidth]{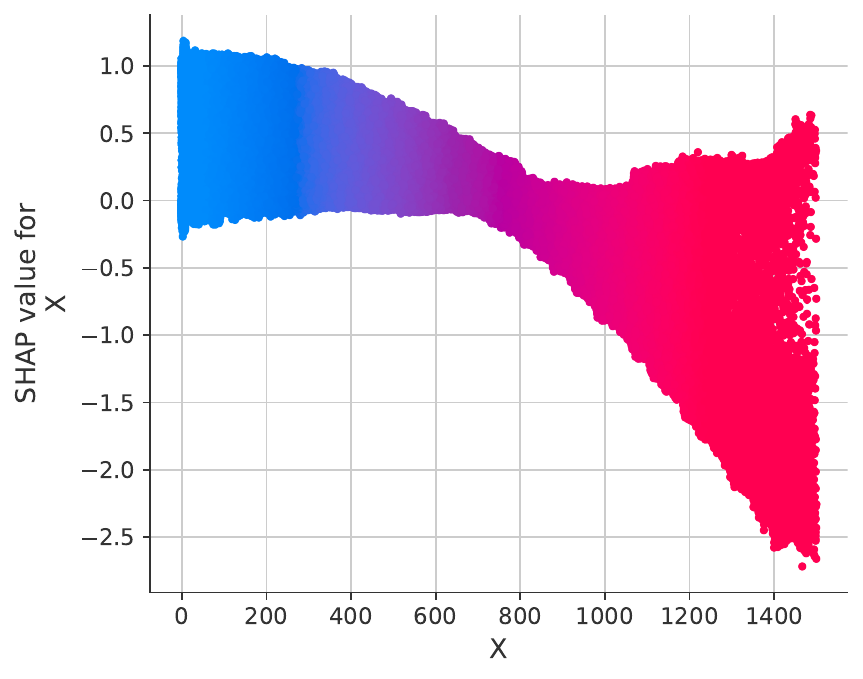}
         \caption{X (cm)}
         \label{fig:x}
     \end{subfigure}
     \hfill
     \begin{subfigure}[b]{0.3\textwidth}
         \centering
         \includegraphics[width=\textwidth]{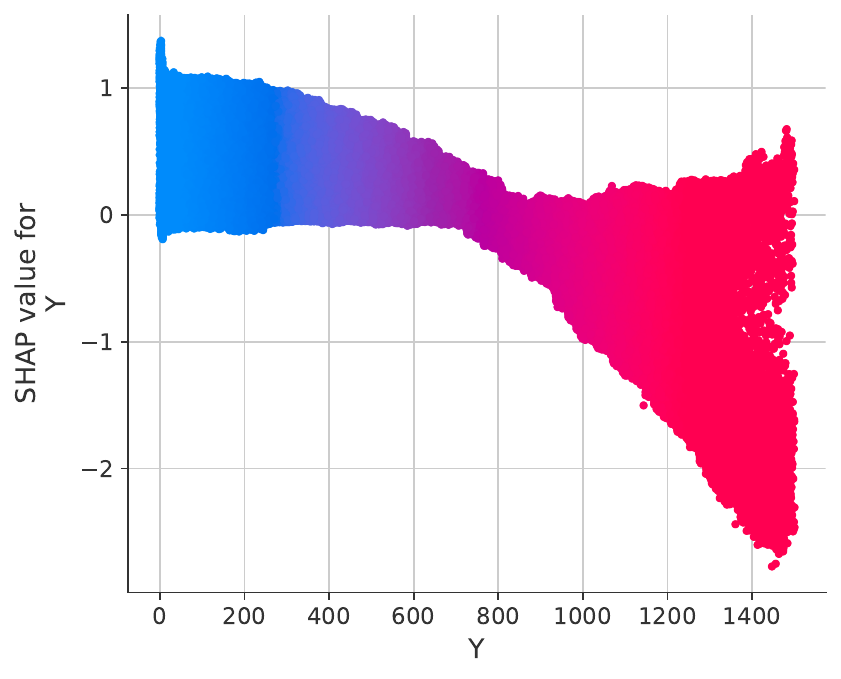}
         \caption{Y (cm)}
         \label{fig:y}
     \end{subfigure}
     \hfill
     \begin{subfigure}[b]{0.3\textwidth}
         \centering
         \includegraphics[width=\textwidth]{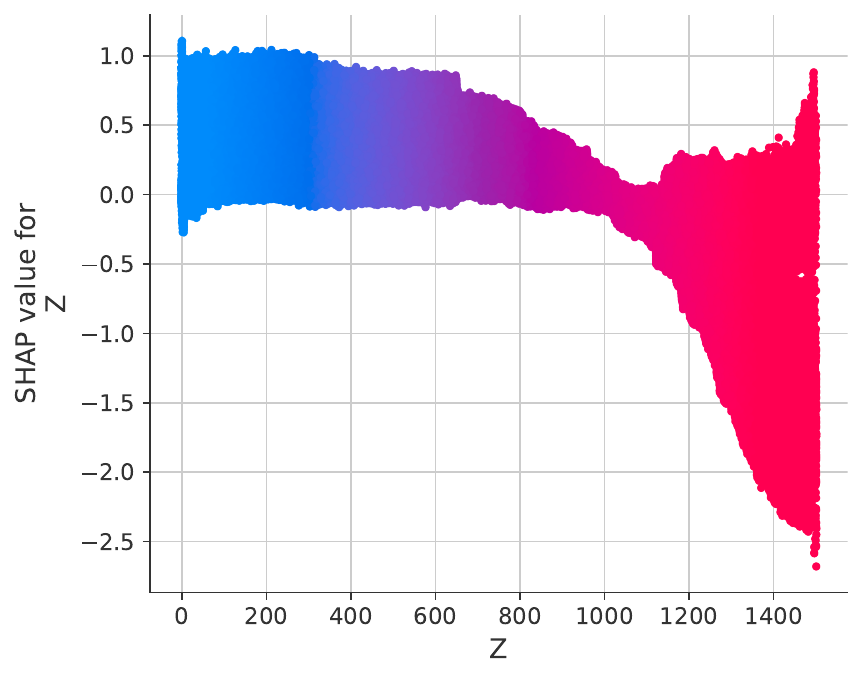}
         \caption{Z (cm)}
         \label{fig:z}
     \end{subfigure}
        \caption{Figure shows the SHAP value for spatial coordinates. \ref{fig:x} shows the X,  \ref{fig:y} shows Y, and \ref{fig:z} shows Z (height of the detector). The color code shows the value of the feature; blue means low values and red means high values.}
        \label{fig:dependency}
\end{figure}

To evaluate the spatial (x,y,z) coordinates, the absolute values of this feature were given to the model. As demonstrated by the figure~\ref{fig:shap values}, higher values of spatial coordinates have an adverse effect on the model. Additionally, the dependency figure ~\ref{fig:dependency} of three spatial coordinates demonstrates that a significant proportion of the higher data points have a negative impact on the model. All three coordinates demonstrate an adverse impact on the model, specifically when the particle is near the detector wall.

The implementation of a spatial coordinate cut enables the enhancement of the model. The optimal range for the variables x and y is defined as the interval [-1000, 1000], whereas the variable z is restricted to the interval [-1200, 1200]. The utilization of this particular cut implementation led to a notable improvement in performance when it came to the prediction of Cherenkov events, reaching an accuracy rate of 98.4\%. Similarly, for scintillation events, the program reached an accuracy rate of 92.9\%. The implementation of a threshold on the variables numPE and charge did not generate a favorable enhancement in the model's performance. Instead, it led to a decline in the model's predictive capacity.


\section{Conclusion}
\label{sec:conclusion}

In this study, we present the simulation and analysis of a detector deployed with a WbLS detection medium. The simulation framework employed in this research is the Rat-Pac simulation program, based on the Geant4 framework. Through this simulation, a comprehensive dataset of Cherenkov and scintillation physics events was generated. 
By using artificial intelligence (AI), we have successfully developed and improved effective machine learning models that are capable of accurately distinguishing between Cherenkov and scintillation signals. Using the XGBoost algorithm, the models were trained on simulated data in order to maximize the separation of Cherenkov and scintillation events. The primary objective was to identify key parameters for identifying these physics events through machine learning techniques.

To assess the performance of our classification models, a comprehensive set of evaluation metrics was employed, including the confusion matrix, accuracy, recall, and f1-score. The precision levels achieved for scintillation and Cherenkov event predictions were notably high, at 83\% and 95\% respectively, as observed on the independent test dataset. Furthermore, a recall rate of 95\% for scintillation and 83\% for Cherenkov events. The f1-score, which harmonizes precision and recall, substantiated the efficacy of our model, particularly in discerning Scintillation events.  The classification performance of our models was evaluated using Receiver Operating Characteristic (ROC) curves, which yielded a measurement of $0.96 \pm 1.2\times 10^{-4}$. This metric showcased the models' robustness and reliability in correctly classifying events.

In a novel addition to this research, we introduced the SHAP (SHapley Additive exPlanations) methodology, rooted in game theory, to unfold the contributions of individual features to the models' final decisions. Through SHAP analysis, we uncovered that the number of hits played a dominant role in model prediction, while mean hit time exhibited the lowest impact. Employing various PMTs with precise time settings can yield varied outcomes in the training dataset. In addition, a higher value of the number of hits and charge demonstrated an adverse effect on model performance, although higher hit time values were shown to have a positive influence.

The findings indicate that detecting Cherenkov light is a more efficient process compared to identifying scintillation light. To enhance the model's capability for scintillation light identification, we introduce a cut on spatial (x,y,z) coordinates. This implementation enhances the accuracy of the model in identifying both Cherenkov and scintillation events. One strategy might be to add a novel feature to the angular parameter that would affect the particle's Cherenkov light's direction. Such an addition could have a valuable effect on the model prediction. Alternatively, considering different design concepts, increasing the number of photomultiplier tubes (PMTs), or adjusting the percentage of WbLS could also be beneficial.

This study helps us better understand using AI and simulation-generated data to distinguish Cherenkov and scintillation light emissions in neutrino detectors. The objective of this study is to establish a foundation for the implementation of Cherenkov/Scintillation detection methods based on artificial intelligence in hybrid neutron detectors.  There is potential for enhancement in this field by carefully considering various factors.

\acknowledgments
This work was supported by the Scientific and Technological Research Council of Türkiye (TÜBİTAK) under project number 123F042. 

The author expresses gratitude to Dr. Vincent Fisher for his valuable contribution to the simulation process of the dataset. Specifically, Dr. Fisher's insights on producing Cherenkov and Scintillation light separately within the program were instrumental. Additionally, the author acknowledges Prof. Dr. Gabriel Orebi Gann and her research group for their efforts in making the WbLS results publicly available and for their implementation in the Rat-Pac simulation program.


\bibliographystyle{JHEP}
\bibliography{biblio.bib}

\providecommand{\href}[2]{#2}\begingroup\raggedright\begin{thebibliography}{10}

\bibitem{Becker-Szendy:1995qpb}
R.~Becker-Szendy et~al., \emph{{Neutrino measurements with the IMB detector}}, \href{https://doi.org/10.1016/0920-5632(94)00765-N}{\emph{Nucl. Phys. B Proc. Suppl.} {\bfseries 38} (1995) 331}.

\bibitem{Becker-Szendy:1992ufh}
R.~Becker-Szendy et~al., \emph{{IMB-3: A Large water Cherenkov detector for nucleon decay and neutrino interactions}}, \href{https://doi.org/10.1016/0168-9002(93)90998-W}{\emph{Nucl. Instrum. Meth. A} {\bfseries 324} (1993) 363}.

\bibitem{FUKUDA2003418}
S.~Fukuda, Y.~Fukuda, T.~Hayakawa and et~al., \emph{The super-kamiokande detector}, \href{https://doi.org/https://doi.org/10.1016/S0168-9002(03)00425-X}{\emph{Nucl. Instrum. Meth. A} {\bfseries 501} (2003) 418}.

\bibitem{BOGER2000172}
J.~Boger, R.~Hahn, J.~Rowley and et~al., \emph{The sudbury neutrino observatory}, \href{https://doi.org/https://doi.org/10.1016/S0168-9002(99)01469-2}{\emph{Nucl. Instrum. Meth. A} {\bfseries 449} (2000) 172}.

\bibitem{PhysRevLett.90.021802}
{\scshape KamLAND Collaboration} collaboration, \emph{First results from kamland: Evidence for reactor antineutrino disappearance}, \href{https://doi.org/10.1103/PhysRevLett.90.021802}{\emph{Phys. Rev. Lett.} {\bfseries 90} (2003) 021802}.

\bibitem{PhysRevLett.94.081801}
{\scshape KamLAND Collaboration} collaboration, \emph{Measurement of neutrino oscillation with kamland: Evidence of spectral distortion}, \href{https://doi.org/10.1103/PhysRevLett.94.081801}{\emph{Phys. Rev. Lett.} {\bfseries 94} (2005) 081801}.

\bibitem{ALIMONTI2009568}
G.~Alimonti, C.~Arpesella, H.~Back and et~al., \emph{The borexino detector at the laboratori nazionali del gran sasso}, \href{https://doi.org/https://doi.org/10.1016/j.nima.2008.11.076}{\emph{Nucl. Instrum. Meth. A} {\bfseries 600} (2009) 568}.

\bibitem{Apollonio2003331}
M.~Apollonio, A.~Baldini, C.~Bemporad and et~al., \emph{Search for neutrino oscillations on a long base-line at the chooz nuclear power station}, \href{https://doi.org/10.1140/epjc/s2002-01127-9}{\emph{European Physical Journal C} {\bfseries 27} (2003) 331 – 374}.

\bibitem{PhysRevLett.108.191802}
{\scshape RENO Collaboration} collaboration, \emph{Observation of reactor electron antineutrinos disappearance in the reno experiment}, \href{https://doi.org/10.1103/PhysRevLett.108.191802}{\emph{Phys. Rev. Lett.} {\bfseries 108} (2012) 191802}.

\bibitem{An_2013}
A.F.~P., A.~Q., B.J.~Z. and et~al., \emph{Improved measurement of electron antineutrino disappearance at daya bay}, \href{https://doi.org/10.1088/1674-1137/37/1/011001}{\emph{Chinese Physics C} {\bfseries 37} (2013) 011001}.

\bibitem{YEH201151}
M.~Yeh, S.~Hans, W.~Beriguete, R.~Rosero, L.~Hu, R.~Hahn et~al., \emph{A new water-based liquid scintillator and potential applications}, \href{https://doi.org/https://doi.org/10.1016/j.nima.2011.08.040}{\emph{Nucl. Instrum. Meth. A} {\bfseries 660} (2011) 51}.

\bibitem{4336931}
D.R.~Winn and D.~Raftery, \emph{Water-based scintillators for large-scale liquid calorimetry}, \href{https://doi.org/10.1109/TNS.1985.4336931}{\emph{IEEE Transactions on Nuclear Science} {\bfseries 32} (1985) 727}.

\bibitem{Land2020oiz}
B.J.~Land, Z.~Bagdasarian, J.~Caravaca, M.~Smiley, M.~Yeh and G.D.~Orebi~Gann, \emph{{MeV-scale performance of water-based and pure liquid scintillator detectors}}, \href{https://doi.org/10.1103/PhysRevD.103.052004}{\emph{Phys. Rev. D} {\bfseries 103} (2021) 052004} [\href{https://arxiv.org/abs/2007.14999}{{\ttfamily 2007.14999}}].

\bibitem{Zhao2023ydx}
R.~Zhao, L.~Bignell, D.E.~Jaffe, R.~Rosero, M.~Yeh, W.~Wang et~al., \emph{{Performance of a ton-scale water-based liquid scintillator detector}}, \href{https://doi.org/10.1088/1748-0221/19/01/P01003}{\emph{JINST} {\bfseries 19} (2024) P01003} [\href{https://arxiv.org/abs/2312.09293}{{\ttfamily 2312.09293}}].

\bibitem{Bignell_2015}
L.~Bignell, D.~Beznosko, M.~Diwan and et~al., \emph{Characterization and modeling of a water-based liquid scintillator}, \href{https://doi.org/10.1088/1748-0221/10/12/P12009}{\emph{Journal of Instrumentation} {\bfseries 10} (2015) P12009}.

\bibitem{D0MA00055H}
D.R.~Onken, F.~Moretti, J.~Caravaca, M.~Yeh, G.D.~Orebi~Gann and E.D.~Bourret, \emph{Time response of water-based liquid scintillator from x-ray excitation}, \href{https://doi.org/10.1039/D0MA00055H}{\emph{Mater. Adv.} {\bfseries 1} (2020) 71}.

\bibitem{PhysRevC.95.055801}
J.~Caravaca, F.B.~Descamps, B.J.~Land, J.~Wallig, M.~Yeh and G.D.~Orebi~Gann, \emph{Experiment to demonstrate separation of cherenkov and scintillation signals}, \href{https://doi.org/10.1103/PhysRevC.95.055801}{\emph{Phys. Rev. C} {\bfseries 95} (2017) 055801}.

\bibitem{LI2016303}
M.~Li, Z.~Guo, M.~Yeh, Z.~Wang and S.~Chen, \emph{Separation of scintillation and cherenkov lights in linear alkyl benzene}, \href{https://doi.org/https://doi.org/10.1016/j.nima.2016.05.132}{\emph{Nucl. Instrum. Meth. A} {\bfseries 830} (2016) 303}.

\bibitem{Kaptanoglu2022}
T.~Kaptanoglu, E.J.~Callaghan, M.~Yeh and G.D.~Orebi~Gann, \emph{{Cherenkov and scintillation separation in water-based liquid scintillator using an $LAPPD^{TM}$}}, \href{https://doi.org/10.1140/epjc/s10052-022-10087-5}{\emph{The European Physical Journal C} {\bfseries 82} (2022) 169}.

\bibitem{Zsoldos:2022uai}
S.~Zsoldos, \emph{{Theia: an advanced optical neutrino detector}}, \href{https://doi.org/10.22323/1.398.0266}{\emph{PoS} {\bfseries EPS-HEP2021} (2022) 266}.

\bibitem{Anderson_2023}
T.~Anderson, E.~Anderssen, M.~Askins and et~al., \emph{Eos: conceptual design for a demonstrator of hybrid optical detector technology}, \href{https://doi.org/10.1088/1748-0221/18/02/P02009}{\emph{Journal of Instrumentation} {\bfseries 18} (2023) P02009}.

\bibitem{ANNIE:2017nng}
{\scshape ANNIE} collaboration, \emph{{Accelerator Neutrino Neutron Interaction Experiment (ANNIE): Preliminary Results and Physics Phase Proposal}},  \href{https://arxiv.org/abs/1707.08222}{{\ttfamily 1707.08222}}.

\bibitem{ANNIE:2023yny}
{\scshape ANNIE} collaboration, \emph{{Deployment of Water-based Liquid Scintillator in the Accelerator Neutrino Neutron Interaction Experiment}},  \href{https://arxiv.org/abs/2312.09335}{{\ttfamily 2312.09335}}.

\bibitem{Psihas:2020pby}
F.~Psihas, M.~Groh, C.~Tunnell and K.~Warburton, \emph{{A Review on Machine Learning for Neutrino Experiments}}, \href{https://doi.org/10.1142/S0217751X20430058}{\emph{Int. J. Mod. Phys. A} {\bfseries 35} (2020) 2043005} [\href{https://arxiv.org/abs/2008.01242}{{\ttfamily 2008.01242}}].

\bibitem{Aurisano_2016}
A.~Aurisano, A.~Radovic, D.~Rocco and et~al., \emph{A convolutional neural network neutrino event classifier}, \href{https://doi.org/10.1088/1748-0221/11/09/P09001}{\emph{Journal of Instrumentation} {\bfseries 11} (2016) P09001}.

\bibitem{PhysRevD.103.092003}
{\scshape MicroBooNE Collaboration} collaboration, \emph{Convolutional neural network for multiple particle identification in the microboone liquid argon time projection chamber}, \href{https://doi.org/10.1103/PhysRevD.103.092003}{\emph{Phys. Rev. D} {\bfseries 103} (2021) 092003}.

\bibitem{PhysRevD.102.092003}
{\scshape DUNE Collaboration} collaboration, \emph{Neutrino interaction classification with a convolutional neural network in the dune far detector}, \href{https://doi.org/10.1103/PhysRevD.102.092003}{\emph{Phys. Rev. D} {\bfseries 102} (2020) 092003}.

\bibitem{DUNE:2020ypp}
{\scshape DUNE} collaboration, \emph{{Deep Underground Neutrino Experiment (DUNE), Far Detector Technical Design Report, Volume II: DUNE Physics}},  \href{https://arxiv.org/abs/2002.03005}{{\ttfamily 2002.03005}}.

\bibitem{7838264}
E.~Racah, S.~Ko, P.~Sadowski and et~al., \emph{Revealing fundamental physics from the daya bay neutrino experiment using deep neural networks},  in \emph{2016 15th IEEE International Conference on Machine Learning and Applications (ICMLA)}, pp.~892--897, 2016, \href{https://doi.org/10.1109/ICMLA.2016.0160}{DOI}.

\bibitem{Aiello_2020}
S.~Aiello, A.~Albert, S.A.~Garre and et~al., \emph{Event reconstruction for km3net/orca using convolutional neural networks}, \href{https://doi.org/10.1088/1748-0221/15/10/P10005}{\emph{Journal of Instrumentation} {\bfseries 15} (2020) P10005}.

\bibitem{2022.978857}
B.~Jamieson, M.~Stubbs, S.~Ramanna and et~al., \emph{Using machine learning to improve neutron identification in water cherenkov detectors}, \href{https://doi.org/10.3389/fdata.2022.978857}{\emph{Frontiers in Big Data} {\bfseries 5} (2022) }.

\bibitem{ELLER2023168011}
P.~Eller, A.T.~Fienberg, J.~Weldert, G.~Wendel, S.~Böser and D.~Cowen, \emph{A flexible event reconstruction based on machine learning and likelihood principles}, \href{https://doi.org/https://doi.org/10.1016/j.nima.2023.168011}{\emph{Nucl. Instrum. Meth. A} {\bfseries 1048} (2023) 168011}.

\bibitem{GEANT4:2002zbu}
{\scshape GEANT4} collaboration, \emph{{GEANT4--a simulation toolkit}}, \href{https://doi.org/10.1016/S0168-9002(03)01368-8}{\emph{Nucl. Instrum. Meth. A} {\bfseries 506} (2003) 250}.

\bibitem{AMAUDRUZ2019373}
P.-A.~Amaudruz, M.~Batygov, B.~Beltran et~al., \emph{In-situ characterization of the hamamatsu r5912-hqe photomultiplier tubes used in the deap-3600 experiment}, \href{https://doi.org/https://doi.org/10.1016/j.nima.2018.12.058}{\emph{Nucl. Instrumen. Meth. A: t} {\bfseries 922} (2019) 373}.

\bibitem{Caravaca2020}
J.~Caravaca, B.J.~Land, M.~Yeh and G.D.~Orebi~Gann, \emph{Characterization of water-based liquid scintillator for cherenkov and scintillation separation}, \href{https://doi.org/10.1140/epjc/s10052-020-8418-4}{\emph{The European Physical Journal C} {\bfseries 90} (2020) 867}.

\bibitem{Chen:2016:XST:2939672.2939785}
T.~Chen and C.~Guestrin, \emph{{XGBoost}: A scalable tree boosting system},  in \emph{Proceedings of the 22nd ACM SIGKDD International Conference on Knowledge Discovery and Data Mining}, KDD '16, (New York, NY, USA), pp.~785--794, ACM, 2016, \href{https://doi.org/10.1145/2939672.2939785}{DOI}.

\bibitem{10.1007/978-3-030-80421-3_37}
F.~Giannakas, C.~Troussas, A.~Krouska, C.~Sgouropoulou and I.~Voyiatzis, \emph{Xgboost and deep neural network comparison: The case of teams' performance},  in \emph{Intelligent Tutoring Systems}, A.I.~Cristea and C.~Troussas, eds., (Cham), pp.~343--349, Springer International Publishing, 2021.

\bibitem{WINTER20022025}
E.~Winter, \emph{Chapter 53 the shapley value},  vol.~3 of \emph{Handbook of Game Theory with Economic Applications}, pp.~2025--2054, Elsevier (2002), \href{https://doi.org/https://doi.org/10.1016/S1574-0005(02)03016-3}{DOI}.

\bibitem{Shapley1953StochasticG}
L.S.~Shapley, \emph{Stochastic games*}, {\emph{Proceedings of the National Academy of Sciences} {\bfseries 39} (1953) 1095 }.

\end{thebibliography}\endgroup


\end{document}